# A Giant Arc in a ROSAT Detected Cluster of Galaxies*

A.C. Edge[1], H. Böhringer[2], L. Guzzo[3], C.A. Collins[4], D. Neumann[2], G. Chincarini[3,5], S. De Grandi[5], R. Dümmler[6], H. Ebeling[1,2], S. Schindler[2], W. Seitter[6], P. Vettolani[7], U. Briel[2], R. Cruddace[8], R. Gruber[2], H. Gursky[8], G. Hartner[2], H.T. MacGillivray[9], P. Schuecker[6], P. Shaver[10], W. Voges[2], J. Wallin[8], A. Wolter[3], and G. Zamorani[11]

[1] Institute of Astronomy, Madingley Road, Cambridge CB3 0HA, UK
[2] Max-Planck-Institut für extraterrestrische Physik, D-85740 Garching, Germany
[3] Osservatorio Astronomico di Brera, Via Bianchi 46, I-22055 Merate (CO), Italy
[4] School of Chemical and Physical Sciences, Liverpool John-Moores University, Liverpool, U.K.
[5] Dipartimento di Fisica, Università di Milano, Via Celoria 16, I-20133 Milano, Italy
[6] Astronomisches Institut der Universität Münster, Münster, Germany
[7] Istituto di Radioastronomia del CNR, Via Gobetti 101, I-40129 Bologna, Italy
[8] Space Sci. Div., Naval Research Lab., Washington D.C., 20375, USA
[9] Royal Observatory Edinburgh, Blackford Hill, Edinburgh EH9 3HJ, U.K.
[10] ESO, D-85748 Garching, Germany
[11] Osservatorio Astronomico, Via Zamboni 33, I-40137 Bologna, Italy

**Abstract.**
In this letter we report on the serendipitous discovery of a giant arc in the Abell Supplementary cluster S295 during optical follow-up observations of clusters detected in the ROSAT All-Sky Survey. The cluster has a redshift of 0.30 and a velocity dispersion of $\sigma_v \simeq 900$ km s$^{-1}$. This is the first giant arc newly found in a ROSAT Survey cluster, indicating the potential advantage of searching for giant arcs in X-ray selected clusters.

## 1. Introduction

Cluster of galaxies act as giant gravitational lenses (Soucail et al. 1987, Lynds & Petrosian 1986). The critical mass necessary to achieve lensing of a background galaxy by a cluster requires a substantial mass (i.e. optically rich) in a relatively small projected area of the sky (i.e. compact). The most advantageous redshift of a cluster to lens a galaxy at a redshift of 0.7–1.5 (the expected redshift of galaxies 25–26 magnitude for which the surface density is high enough to expect to find a few galaxies behind each cluster) is 0.2–0.5. Deep surveys of distant, optically-selected clusters do find a high frequency of arcs, for instance one per cluster in the sample of Smail et al. (1991). Similarly, X-ray selected samples such as the EMSS (Gioia et al. 1990) also show frequent gravitational arcs (Le Févre et al. 1994, Gioia & Luppino 1994). However, only a small fraction (2 in 19 in the Smail et al. sample and 3 in 41 in the Gioia & Luppino sample) show high amplification, giant arcs such as the ones found in A370 (Soucail et al. 1987) and Cl2244-02 (Hammer et al. 1989). Despite their inherent rarity, giant arcs are of great importance as they offer an opportunity to measure the redshifts of galaxies significantly fainter than present instrumentation allows. Modelling of the shape of a giant arc can tightly constrain the cluster potential as demonstrated by Kneib et al. (1994) in A370. Also the frequency of giant arcs can be used to place statistical limits on the distribution of mass in clusters and the redshift distribution of background galaxies (Wu & Hammer 1993).

In 1992 we commenced an ESO Key Programme aimed to obtain redshifts for all the clusters of galaxies detected by the ROSAT X–ray satellite in the Southern Hemisphere within a certain flux limit. As a by-product of the observing strategy adopted, short CCD images for all clusters are obtained for the preparation of multi–slit masks for the ESO Faint Object Spectrograph and Camera (EFOSC). These images are deep enough to detect giant arcs of the size and surface brightness found in other clusters. With a flux limit of 0.1 ROSAT counts sec$^{-1}$ (equivalent to an unabsorbed flux of 1.2–1.4×10$^{-12}$ ergs s$^{-1}$ cm$^{-2}$ in the 0.1–2.5 keV band), we expect to observe up to fifty distant, X-ray luminous clusters, which will provide a limit on the frequency of giant arcs in the most massive clusters. In this paper we present a giant arc serendipitously discovered in the second observing run of our Key Programme, making it the first giant arc found in a new ROSAT-selected cluster. It was detected in the Abell Supplementary cluster S295 which is a Richness Class 2, Distance Class 6 cluster. We present R and B images of the cluster, the cluster redshift and an estimate of the velocity dispersion based on 6 galaxy redshifts.

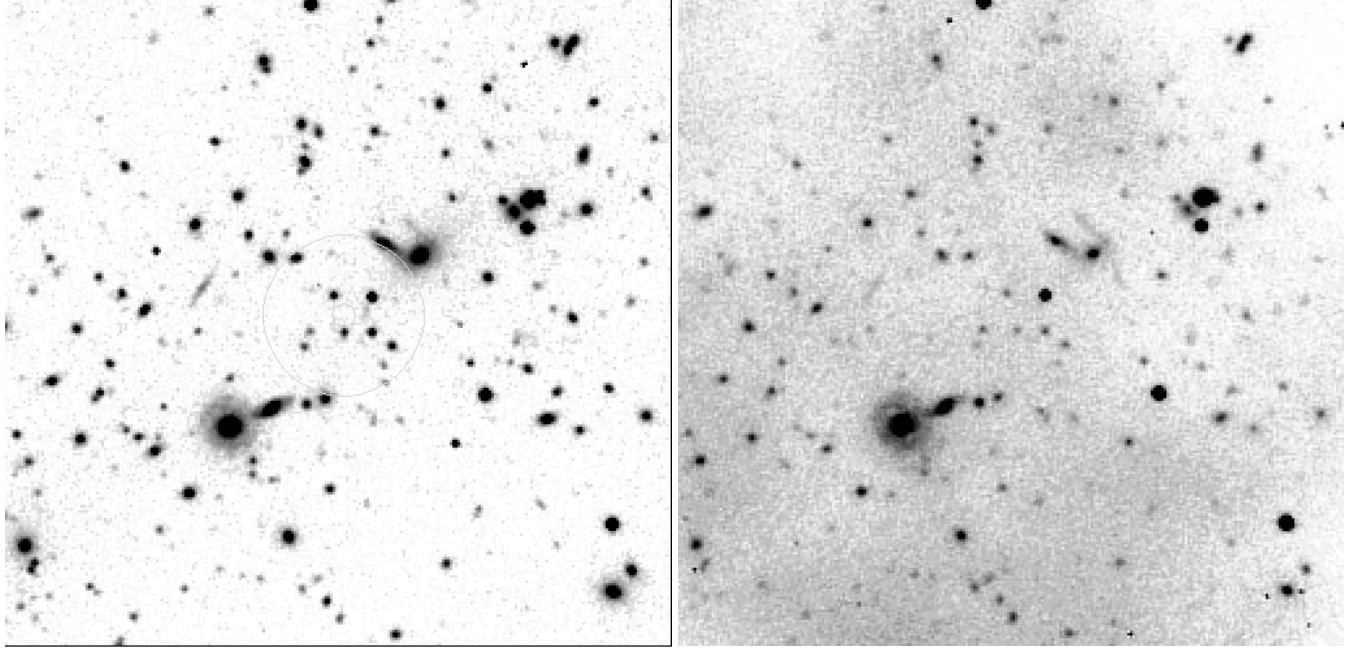

**Fig. 1.** a & b. R and B images of S295 with exposures of 120s and 600s respectively over a 2.6 arcminute square field centred on the X-ray position. The circles on the R image mark the X-ray position with its error. Both images are rotated to have North up and East left.

## 2. Results

### 2.1. Optical Images

The images presented in Figs. 1 a & b were taken using EFOSC on the ESO 3.6 m telescope during the night of 27/28[th] November 1992 in relatively good seeing (1.1–1.2″). Fig. 1a is the discovery frame taken in R with an exposure of 120s. Fig. 1b is a follow-up frame taken in B with an exposure of 600s shown to highlight the differences in colour between the arc and the cluster galaxies.

In both frames a clear arc can be seen North-West of the galaxy concentration. The arc is most obvious in the B image, as the brightest cluster member is within 4″ of the arc and the contrast is reduced in the R frame. Although neither frame is photometrically calibrated, we estimate the colour of the arc to be $B - R \approx 1.5$ with a surface brightness of approximately 25 mag arcsec$^{-2}$ in B, consistent with other bright arcs (Soucail 1991). The arc is 20 arcsec long, not resolved in width and has an apparent radius of curvature of 15–22″. There are three 'knots' in the arc, one at either end and one in the middle. The exposure times for both CCD images are not sufficient to provide a more detailed morphology of the arc. Fortunately detailed follow-up has been performed by the Toulouse group, following our communication. From deep NTT images the distortions visible at both ends in Figs. 1 a & b show up as a 'Mexican Hat', indicating a complex potential shape. The redshift of the arc, 0.93 (Fort & Mellier 1994), is consistent with the gravitational lens origin and its colour. There is a second 'arc-like' structure to the North-East of the X-ray position, which is very likely to be an edge-on spiral, as a bulge can be seen in the R image, and not a gravitational arc.

The image shows in excess of 25 galaxies within a 0.5 $h^{-1}$ Mpc radius from the X-ray position, between the magnitude of the third brightest galaxy, $m_3$, and $m_3 + 2$, i.e. the Bahcall number (Bahcall 1977). This number is uncertain as it is not possible to quantify the contribution from foreground galaxies. However, a Bahcall number of 30 is consistent with that found for similarly X-ray luminous clusters at lower redshift (Edge & Stewart 1991a). From the galaxy redshifts obtained the following night, the brightest galaxy (a face-on spiral to the South-East of the X-ray position) was found to be a foreground object ($z = 0.0969$), while from 6 cluster galaxies we obtained a mean redshift $z_{cl} = 0.3007$, with a velocity dispersion $\sigma_v \simeq 907$ km s$^{-1}$. This high value for $\sigma_v$, although tentative, is consistent with the picture suggested by the strong X-ray emission, high galaxy density and the gravitational lensing.

### 2.2. X-ray properties

The cluster was detected in the ROSAT All Sky Survey (RASS) with a count rate of 0.114 ± 0.016 counts s$^{-1}$ in the 0.5 to 2.0 keV band. The X-ray spectrum is consistent with emission from hot intracluster gas and the X-ray emission is significantly extended. These two points indicate that the cluster is the optical counterpart to the source. The X-ray emission is centred on a position of 02 45 27.0 -53 02 00.0 (2000) with a 20″ error radius. This position is 24″ from the central galaxy which is consistent with the offsets seen for other clusters in the RASS. From the 75 detected photons an apparent elongation along PA 36° (north to east) is found.

Using the observed X-ray luminosity and gas temperature relation (Edge & Stewart 1991b; David et al. 1993) as a guide, we derive an X-ray luminosity of $1.09 \pm 0.18 \times 10^{45}$ erg s$^{-1}$ (0.1–2.4 keV, unabsorbed) for a range of temperatures between 4–12 keV and the interstellar column density of $N_H = 4 \times 10^{20}$ cm$^{-2}$ (Dickey & Lockman 1990). The corresponding unabsorbed, bolometric luminosity is $2.0^{+0.7}_{-0.5} \times 10^{45}$ erg s$^{-1}$.

a radius of between 6–25″ (either the position of the cD or the X-ray position), we estimate a projected mass between $4\times10^{12}$ M$_\odot$ at 30 kpc and $5\times10^{13}$ M$_\odot$ at 100 kpc. This is comparable to other lensing mass estimates in the cores of X-ray luminous clusters (Mathez et al. 1992; Babul & Miralda-Escudé 1994). A high resolution X-ray image of the cluster is vital to reducing the uncertainty in the centre of the cluster and hence the mass.

## 3. Conclusions

We have discovered a giant arc in an optically unremarkable Abell Supplementary cluster which was selected from the RASS for its high X-ray flux. The arc is similar in colour and surface brightness to giant arcs found in other clusters. Given the constraints of the observing strategy adopted by a such large observing programme and the single detection of an arc, it is not possible to put any interesting limits on the frequency of arcs in ROSAT-selected clusters at this stage, although it will be when the programme is complete.

This result shows that, even for X-ray selected samples, the chance of a finding a prominent arc in a specific cluster is small, so bright arcs do not offer a systematic technique to study of individual cluster masses. On the other hand, these bright arcs do offer a direct opportunity to determine redshifts of galaxies well beyond present limits. Only with substantially deeper optical imaging and high resolution X-ray images will it be possible to determine the cluster mass distribution on a cluster by cluster basis (Smail et al. 1994).

*Acknowledgements.* We thank the supporting staff of the European Southern Observatory for their excellent assistance, in particular J. Peres for his continuous help with the PUMA machine. The exceptional properties of this cluster would not have been uncovered without the immense effort of the ROSAT team and we gratefully acknowledge all those at MPE. ACE thanks Prof. Trümper for the hospitality shown on the many visits to MPE over the last 4 years, and SERC for support. This work has received partial financial support from EEC contract Nr. ERB–CHRX–CT92–0033.